\documentclass[review]{elsarticle}

\usepackage{lineno,hyperref,graphicx}
\modulolinenumbers[5]

\journal{Acta Astronautica}

\begin{document}

\begin{frontmatter}

\title{SETI in Russia, USSR and the post-Soviet space: \\ a century of research}


\author[GAISH-address]{Lev M. Gindilis}

\author[JIVE-address,TUD-address]{Leonid I. Gurvits\corref{mycorrespondingauthor}}
\cortext[mycorrespondingauthor]{Corresponding author, lgurvits@jive.eu}

\address[GAISH-address]{Sternberg Astronomical Institute, Universitetskiy av. 13, Moscow 119992, Russia}
\address[JIVE-address]{Joint Institute for VLBI ERIC, Oude Hoogeveensedijk 4, \\ 7991 PD Dwingeloo, The Netherlands}
\address[TUD-address]{Department of Astrodynamics and Space Missions, Delft University of Technology, \\
Kluyverweg 1, 2629 HS Delft, The Netherlands}

\begin{abstract}
Studies on extraterrestrial civilisations in Russia date back to the end of the 19th century. The modern period of SETI studies began in the USSR in the early 1960s. The first edition of the I.S. Shklovsky's book {\it Universe, Life, Intelligence} published in 1962 was a founding stone of SETI research in the USSR.  A number of observational projects in radio and optical domains were conducted in the 1960s--1990s. Theoretical studies focused on defining optimal spectral domains for search of artificial electromagnetic signals, selection of celestial targets in search for ETI, optimal methods for encoding and decoding of interstellar messages, estimating the magnitude of astro-engineering activity of ETI, and developing philosophical background of the SETI problem.
Later, in the 1990s and in the first two decades of the 21st century, in spite of acute underfunding and other problems facing the scientific community in Russia and other countries of the former Soviet Union, SETI-oriented research continued. In particular, SETI collaborations conducted a number of surveys of Sun-like stars in the Milky Way, searched for Dyson spheres and artificial optical signals. Several space broadcasting programs were conducted too, including a radio transmission toward selected stars. Serious rethinking was given to incentives for passive and active participation of space civilisations in SETI and CETI.
This paper gives an overview of past SETI activities. It also gives a comprehensive list of publications by authors from Russia, the Soviet Union and the post-Soviet space, as well as some SETI publications by other authors. The rich heritage of SETI research presented in the paper might offer a potentially useful background and starting point for developing strategy and specific research programs of the near future. 
\end{abstract}


\end{frontmatter}


\section{Introduction}

Search for life and intelligence outside of Earth is one of the central scientific philosophical issues throughout the history of the terrestrial civilisation. The start of the modern era in the search for extraterrestrial intelligence (SETI) is usually attributed to the pivotal paper by G.~Cocconi and P.~Morrison in {\it Nature} journal in 1959 \cite{CoMo59}. This played a ``trigger'' role for SETI research around the world. Soviet scientists, primarily astronomers, were strongly influenced by this paper too. However, there was an additional stimulus for SETI in the USSR in the end of the 1950s through the beginning of the 1960s: the advent of the space exploration era. The early world-leading space exploration program was the most remarkable positive achievement of the USSR in the post-World War II period. Not surprisingly, in those years, anything related to space used to attract a lot of attention in the USSR among scientists and general public alike. The SETI problem fit this space-related interest very naturally. It is worth noticing that the first edition of the book {\it Universe, Life, Intelligence} by I.S.~Shklovsky \cite{ISSh62} was prepared for the fith anniversary of the Sputnik launch celebrated in 1962, and the publication was enthusiastically supported by the then President of the Soviet Academy of Sciences and the ``Chief Theorist of cosmonautics''\footnote{The title ``Chief Theorist of cosmonautics'' was used in Soviet mass media as a match to the title ``Chief Designer of cosmonautics'' (S.P.~Korolev). The titles were introduced in order to express the highest respect to the two leading figures of the Soviet space programme. However, these titles were not accompanied with names in any contemporary publications, and the real identities of these two leaders were not disclosed until their deaths.}, M.V.~Keldysh. 

\begin{figure}[t]
   \centering
\includegraphics[width=40.5mm,angle=0]{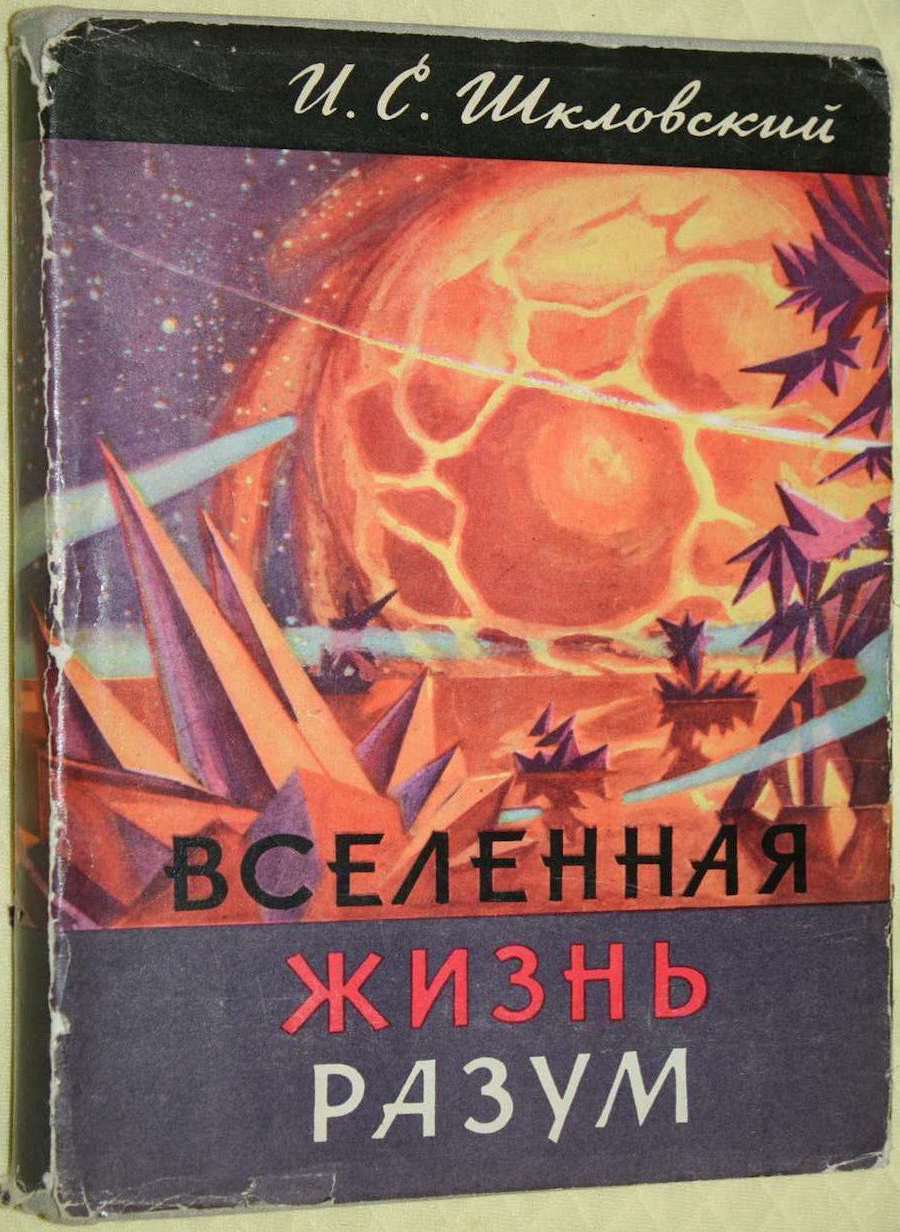}
\includegraphics[width=79.5mm,angle=0]{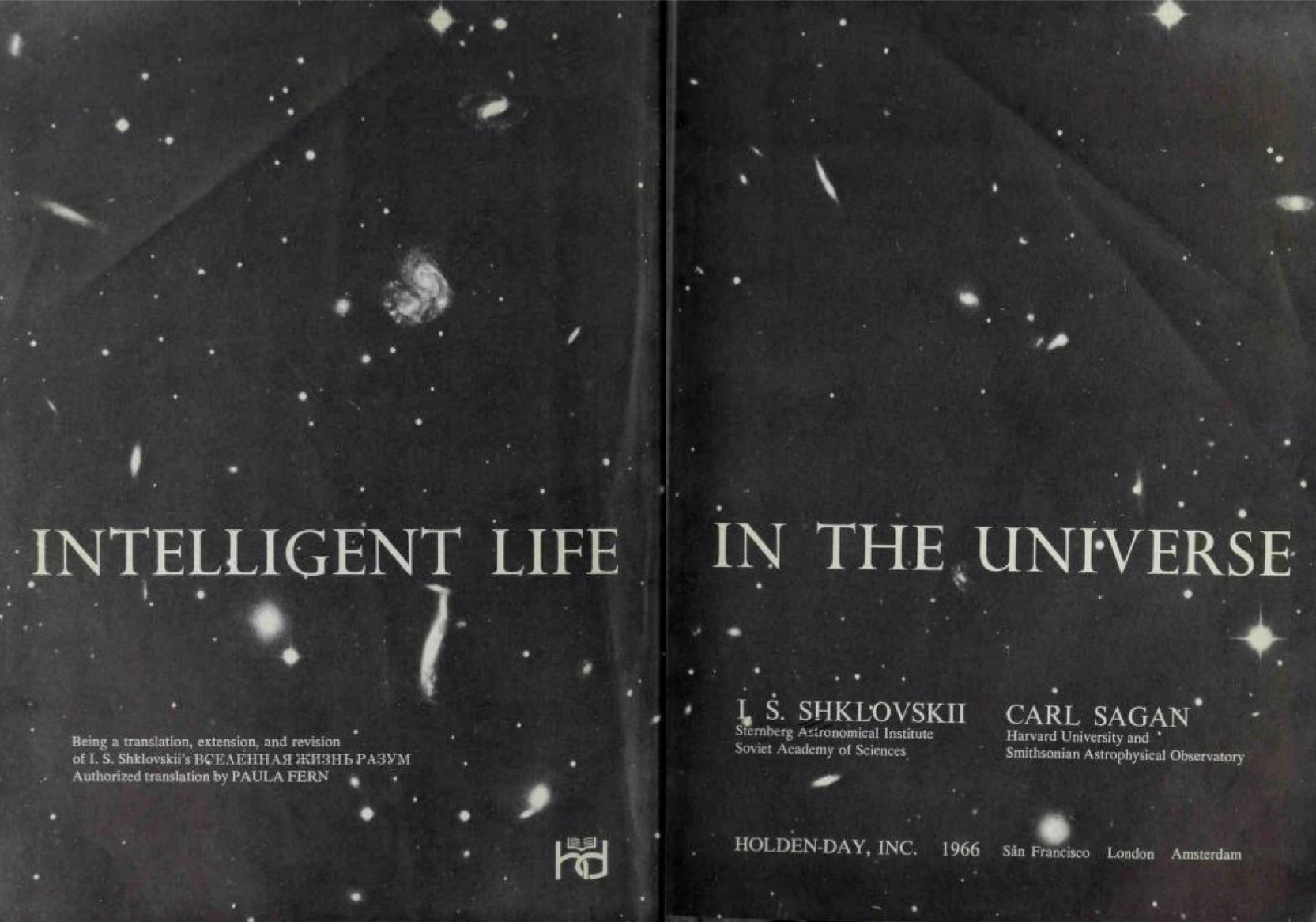}
\caption{The covers of: ({\it left panel}) the first edition (1962) of the book by I.S. Shklovsky {\it Universe, Life, Intelligence} \cite{ISSh62};  ({\it right panel}) the first edition (1966) of the book by I.S.~Shklovskii and C.~Sagan {\it Intelligent Life in the Universe} \cite{ShklSagan66}.}
   \label{The_books}   
\end{figure}

Over the following decades, the SETI problem attracted attention by many researchers in the USSR. Their work resulted in the formulation of important scientific results and generated many ideas, some of which might still remain under-demanded. While some of these studies resulted in publications accessible to the world-wide community of SETI experts, a significant fraction of the body of relevant scientific work remained either behind the ``iron curtain'' or beyond reach due to the linguistic barrier. An interesting illustration of the difficulties that faced Soviet researchers in making their work known outside of the country is the story of Shklovsky's book {\it Universe, Life, Intelligence}. A rumour about this book spread outside the Soviet Union quickly after its publication in 1962. The author got offers for translation and publication abroad. However, restrictive administrative hurdles were too high. At this point, Carl Sagan, who got very much interested in publishing this book in English, and Iosif Shklovsky invented an ingenious solution: let the English book be co-authored by both Shklovsky and Sagan. As a joint work of two co-authors, the book was no longer a subject of severe Soviet restrictions on direct translations. Indeed, a co-authored book \cite{ShklSagan66} that included a translation of the original book by Shklovsky but amended with many new paragraphs written by Sagan was published in 1966 (Fig.~\ref{The_books}). Showing impeccable tact, Sagan, who controlled preparation of the book's publication in the United States, insisted that his additions be clearly highlighted in the American edition.

This paper represents an attempt to provide if not a full anthology of SETI work in Russia, the USSR and the post-Soviet space but at least a major relevant bibliography. One of the authors published similar reviews in Russian \cite{Gin86,Gin88,Gin04}. But even the most recent of these was published 15 years ago. The present paper is based on but not limited to a presentation given at the SETI session of the 40th COSPAR Scientific Assembly in Moscow in 2014. While we cannot predict when SETI will result in a trustworthy detection, in 10, 100, 1000 years if ever at all, a look-back into a wide range of SETI ideas might help in addressing the outcome of future observing programs. Our motivation for the current paper was to add a large set of potentially useful references into the global pull of SETI publications.

The paper's sections \ref{Early-hist}-\ref{Ru-1990s} follow roughly the chronology of SETI studies. However, at various places, the chronological order is violated in favour of topical consistency. In the post-Soviet era, SETI research continues in Russia and Ukraine. These efforts are presented in sections \ref{Ru-1990s} and \ref{Ukr-1990s}, respectively. In lieu of conclusions section \ref{Cnclsns} gives a brief outlook of the near future SETI research.

\section{Early history of SETI in Russia}
\label{Early-hist}

It is hard to define when the interest in Russia to the problem of the existence of intelligent life in the Universe was born. Probably it developed along the general direction of European scientific and philosophical thought. There was a surge of interest to this problem at the end of 19th century. In 1875, a book by E.E.~Neovius, a Russian scientist of Finnish and Swedish heritage, entitled {\it The Most Important Task of Our Time} was published \cite{Neo1875}. In this book he clearly formulated the task of establishing contact with extraterrestrial civilisations and proposed a specific project of communication with inhabitants of the Solar System planets using light signals. He also proposed a language for these communications based on mathematical logic which would permit messages of increasing complexity. A great contribution to the understanding of the problem was made by Russian philosophers of the ``silver age'' (late 19th - early 20th centuries). Some of them stood on science-based positions, others adopted religious-defined approaches. K.E.~Tsiolkovsky played an especially important role. He developed many ideas on space flight \cite{Ts14}. In the middle of the 20th century, G.A.~Tikhov conducted a series of studies on astro-botany, primarily having in mind the possibility of finding life on Mars \cite{Ti49}. These works caused a sharp discussion in the scientific community. One of the Tikhov's main opponents was V.G.~Fesenkov. In 1956, V.G.~Fesenkov and A.I.~Oparin  published a book {\it Life in the Universe} \cite{Opa56}.

\section{Beginning of the modern phase}
\label{Beginng}

In 1959, {\it Nature} published an article by G.~Cocconi and P.~Morrison on the possibility of radio communication with extraterrestrial civilisations \cite{CoMo59}. In 1960, the first searches of signals from extraterrestrial civilisations at the wavelength of 21~cm were carried out at the National Radio Astronomy Observatory (NRAO) in Green Bank (WVA, USA) by Frank Drake under the leadership of Otto Struve \cite{Dr60}. In the 1960--1961, several serious works on possible ways of searching for extraterrestrial civilisations were published by prominent Western radio astronomers R.~ Bracewell \cite{Br60}, F.~Drake \cite{Drake60} and S.~von~Hoerner \cite{Ho61}.

In the USSR, the first to respond to all of these exciting events was Iosif Samuilovich Shklovsky. In 1960, he published a large article entitled ``Is it possible to communicate with intelligent beings of other planets'' in the scientific-popular magazine {\it Priroda} ({\it Nature} in Russian)  \cite{ISSh60}, which formed the basis of his book {\it Universe, Life, Intelligence}, published in 1962 \cite{ISSh62} and then re-published in Russian six times (the latest seventh enlarged edition of 2006, \cite{ISSh06}). Shklovsky attracted to the study of this problem his former student Nikolai Kardashev. In 1964, Kardashev published a paper titled ``Transmission of information by extraterrestrial civilisations'' in the main Soviet astronomical journal, {\it Astronomicheskii Zhurnal}  \cite{NSK64}, in which the idea of the existence of three types of civilisations was formulated. This paper also discussed parameters of continuous isotropic broadband transmission from highly developed civilisations, addressed the question of the optimum wavelength bands for interstellar communication, formulated criteria of artificiality of the signals, and presented possible spectra of the artificial signals. In this paper, Kardashev also noted explicitly two celestial radio sources of an unknown nature, CTA~21 and CTA~102, discovered recently by the Caltech group, which looked unusual due to their compactness and peaked radio spectra with the maxima about 1~GHz. This peak frequency was close to the optimum suggested by Kardashev for radio communication between cosmic civilisations. The English version of this paper \cite{NSK64} has a serious translation error: while the original Russian version states that ``... CTA~21 and CTA~102 have angular sizes less than 20 seconds of arc ...'', the English version reads ``...[these sources] display angular dimension not less than 20 seconds of arc''. This is a critical difference in the SETI context as well as for the interpretation of observations of these sources discussed in section \ref{Frst-stps}. The book by I.S.~Shklovsky and the paper by N.S.~Kardashev\footnote{The paper by N.S. Kardashev \cite{NSK64} had a visible immediate impact to such an extent that even New York Times carried its detailed narration on 29 Nov. 1964} set the stage for SETI studies in the USSR.

\section{The first steps, networking and brainstorming}
\label{Frst-stps}

In May 1964, the First All-Union Conference on Extraterrestrial Civilisations was held at the Byurakan Astrophysical Observatory in Armenia. Major ideas on the ``radio-communicational'' strategy of SETI  were presented by N.S.~Kardashev, V.S.~Troitsky, V.A.~Kotel'nikov, V.I.~Siforov, S.E.~Khaikin and others. They formed the basis for the approaches for SETI research for many subsequent years. Proceedings of the conference were published in Yerevan in 1965  in Russian \cite{ETIr64} and in Jerusalem in 1967 -- in English \cite{ETIe64}. Following a resolution adopted at the Byurakan Conference, the Council on Radio Astronomy of the USSR Academy of Sciences formed a section named ``Search for Extraterrestrial Civilisations''. V.S.~Troitsky was approved by the Council as the Chairman of the Section, N.S.~Kardashev -- as a Vice-chairman, and L.M.~Gindilis -- as a Scientific Secretary. One of the first tasks of the Section was a development of a research program on the problem of communications with extraterrestrial intelligence (CETI). This work was carried out in several stages. In 1974, this program was presented first in a small print brochure, and then published  in {\it Astronomicheskii Zhurnal} in the same year \cite{ProCETI}.

\begin{figure}[t]
    \centering
\includegraphics[width=\textwidth]{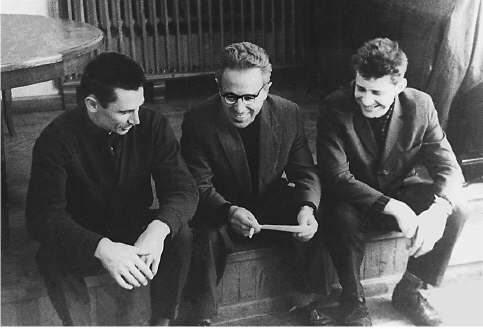}
\caption{G.B.~Sholomitsky, I.S.~Shklovsky and N.S.~Kardashev (left to right) after the press-conference on the discovery of variability of the radio source CTA~102 in the conference-hall of the Sternberg Astronomical Institute, Moscow, 13 April 1965.}
    \label{fig-GBS-ISS-NSK}
\end{figure}

Immediately after the Byurakan Conference of 1964, two groups of Soviet scientists began preparing experiments of searching for radio signals from extraterrestrial civilisations. In 1968-1969, the group led by V.S.~Troitsky in Gorky (now - Nizhnii Novgorod) searched for signals from the nearest solar-type stars (11 stars) and the galaxy M~31 \cite{Tr71}. The second group led by N.S.~Kardashev in Moscow was engaged in research of peculiar sources possessing signs of artificiality and preparation for sky surveys at the centimeter wavelengths, which were considered to be optimal. The list of targets, among others, included two peculiar sources, CTA~21 and CTA~102, identified by Kardashev \cite{NSK64} as requiring special attention. This attention did pay off: one of the sources, CTA~102, demonstrated variability, a clear indication of its compact linear size, not exceeding the light travel distance over the characteristic time of radio variability \cite{Sho65}. Since variability was assumed by Kardashev as an indication on artificiality of the source, this work attracted a lot of attention from the national and world media. A special press-conference triggered by a telegram of the Telegraph Agency of Soviet Union (TASS) \cite{TASS65} and attended by dozens of correspondents from major world news media was held at the Sternberg Astronomical Institute on 13 April 1965 (Fig.~\ref{fig-GBS-ISS-NSK}). he next day, the major Soviet daily {\it Pravda} carried an editorial drafted by I.S.~Shklovsky with a brief balanced summary of the Sholomitsky's  discovery.\footnote{An interesting angle on the story of the CTA~102 variability is presented by a journalist A.P.~Midler, the author of the original TASS telegram, in his article in the second edition of the book of memoirs about I.S.~Shkovsky {\it Intelligence, Life, Universe} to be published in Russian in 2019.}  In the immediate aftermath of this announcement, a number of observatories around the world attempted and failed to confirm the variability of CTA-102. It was not until 1972 when the variability of radio emission of CTA-102 was confirmed by R.W.~Hunstead \cite{Hun72}. While the hypothesis on the artificiality of the CTA-102 radio signal wound down quickly, the discovery of radio variability of quasars made a fundamental impact on the understanding of the nature of this class of objects and triggered developments of high-resolution radio astronomy techniques, in particular -- the Very Long Baseline Interferometry (VLBI).  

Another outcome of the Byurakan Conference was the work at the Sternberg Astronomical Institute on a project of a Kraus-type radio telescope RT-MSU with the shortest observing wavelength of 4~mm. Later, a combination of this project with the APP project developed at the Pulkovo Observatory in Leningrad (now -- St.Petersburg) led to the concept of the RATAN-600 radio telescope built in the North Caucasus. The search for ETI signals was one of the design tasks of RATAN-600.

A boost in interest to the SETI problem was ignited by the  book ''Extraterrestrial civilisations. Problems of interstellar communication", published in Russian in 1969 \cite{ETIr69}, and then translated into English \cite{ETIe71} and Czech \cite{ETIc72}.

\begin{figure}[t]
    \centering
\includegraphics[width=\textwidth]{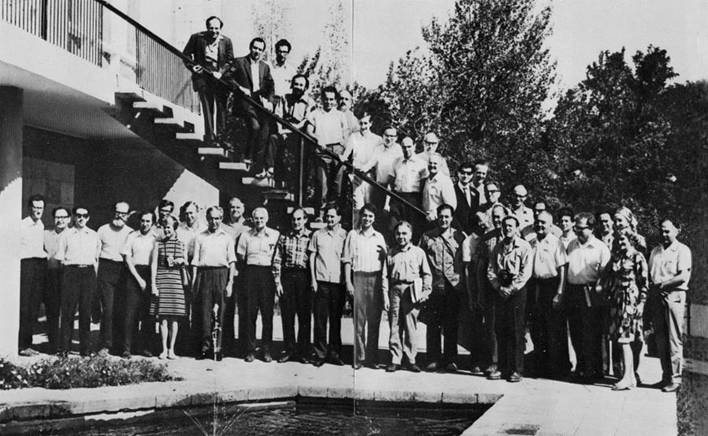}
\caption{Participants of the First US--USSR CETI Conference, Byurakan Astrophysical Observatory, Armenia, September 1971.}
    \label{CETI-71}
\end{figure}

In  September 1971, the Byurakan Astrophysical Observatory hosted the First US––USSR Conference on CETI. In spite of its ``bilateral'' origin, de facto the conference attendance was wider than those two countries alone. This conference provided a stage for pivotal discussions that shaped up the directions of SETI research for several decades. It was attended by two Nobel Prize winners, F.~Crick and Ch.H.~Townes. One participant, V.L.~Ginzburg, would become a Nobel Prize Laureat in 2003. A number of leaders of world astronomical, space and SETI research attended too: V.A.~Ambartsumian, S.Ya.~Braude, B.F.~Burke, F.D.~Drake, F.J.~Dyson, S.~von~Hoerner, S.A.~Kaplan, N.S.~Kardashev, K.I.~Kellermann, G.~Marx, P.~Morrison, B.M.~Oliver, R.~Pe\v{s}ek, C.E.~Sagan, Yu.N.~Pariiskii, I.S.~Shkovsky, V.S.~Troitsky and others (Fig.~\ref{CETI-71}). The proceedings of this conference were published in English in 1973 \cite{CETI73} and in Russian in 1975 \cite{CETI75}. These books remained a desk-top reference for many SETI researchers around the world for several decades.

\subsection{SETI networking and brainstorming}

\begin{figure}[t]
    \centering
\includegraphics[width=\textwidth]{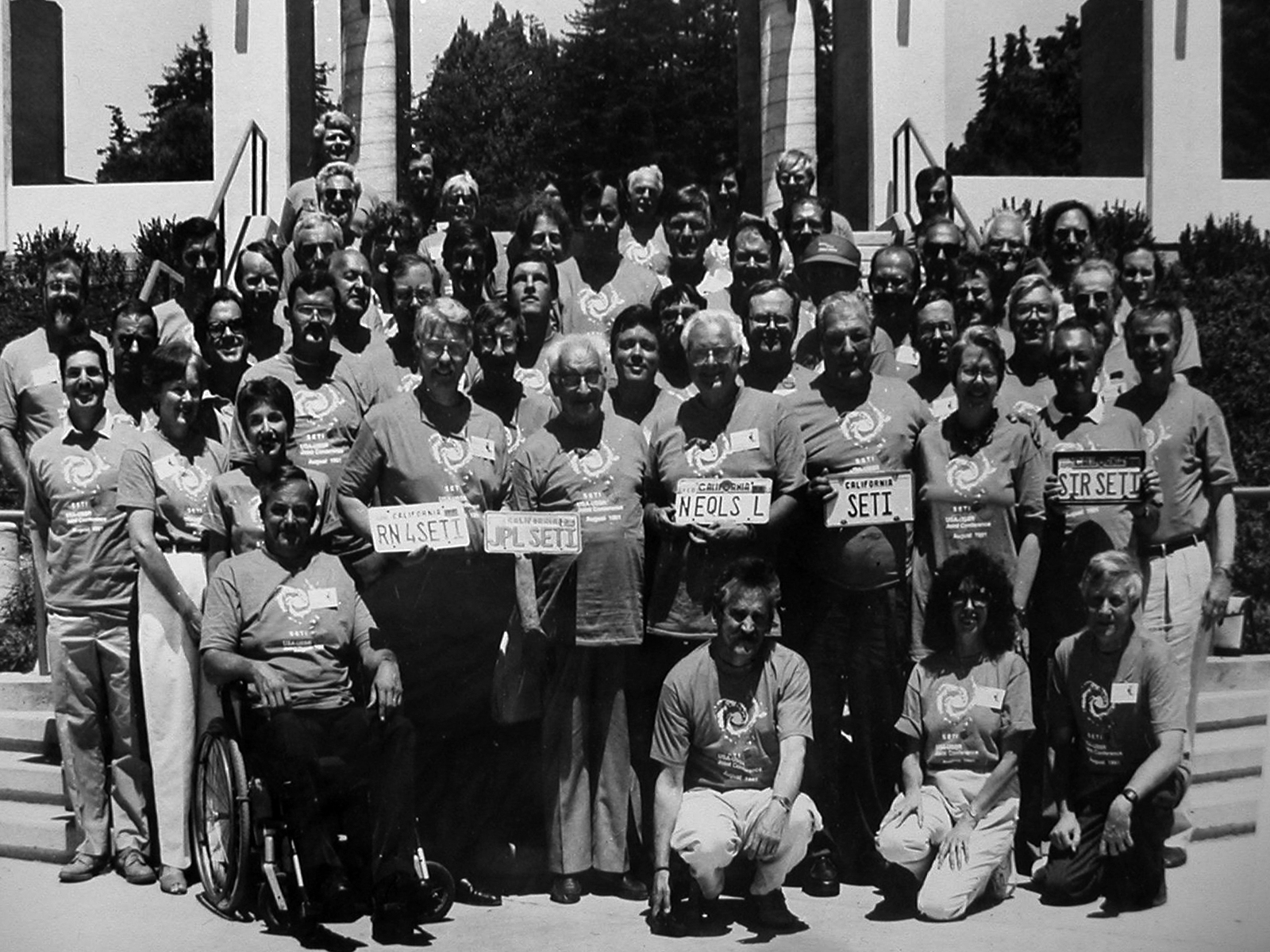}
\caption{Participants of the third Decadal US--Soviet Conference on SETI (Santa Cruz, California, USA, 1991.}
    \label{Santa-Cruz}
\end{figure}

The two Byurakan ``firsts'', the first All-Union Conference of 1964 and the First US--USSR Conference of 1971 have proven clearly the importance of extensive exchange of opinions between SETI researchers working in a broad range of science domains. A number of follow-up conferences, workshops and schools were organised in the USSR and the republics of the former Soviet Union over the past half a century. In  addition to the two Byurakan meetings of 1964 and 1971, the most noticeable milestones of this period of the SETI research in the USSR and post-Soviet space were the Zelenchuk School--Workshop (1975), Tallinn (1981) and Vilnius (1987) Symposia, the third Decadal US--Soviet Conference on SETI (Santa Cruz, California, 1991, Fig.~\ref{Santa-Cruz}), Conference ``Horizons of astronomy and SETI'' (Nizhnii Arkhyz, Russia, 2005), the 3rd IAA Symposium on Searching for Life Signatures (St. Petersburg, Russia, 2011). It is difficult to acknowledge the many organisers of these events who made them a success.  The generous hosts and enthusiastic organisers of these events were V.A.~Ambartsumian (Byurakan, Armenia, 1964, 1971), V.F.~Shvartsman (Pastukhov Mountain, BTA, Russia, 1975), K.~Rebane (Tallinn, Estonia, 1981), V.~Strai\v{z}is and G.~Kakaras (Vilnius, Lithuania, 1987),Yu.Yu.~Balega (Nizhnii Arkhyz, Russia, 2005), A.M.~Finkelstein (St. Petersburg, Russia, 2011). Their leading facilitating role is unquestionable. A major coordinating role has been fulfilled by the SETI Scientific--Cultural Center established in Moscow, Russia in 1992 \cite{Gin93}.

\section{Search for signals and related studies}
\label{Sgnls}

\subsection{Search for impulse callsigns with non-directional antennas}

It was assumed that powerful short broadband pulses can be used as callsigns for interstellar communication. The search for such signals was conducted by two groups: the group in Gorky (now Nizhny Novgorod) at the Radio Physical Research Institute (NIRFI) led by V.S.~Troitsky, and the Moscow group at the Sternberg Astronomical Institute (GAISh) and Space Research Institute (IKI) led by N.S.~Kardashev. At NIRFI, the search was carried out at wavelengths of 50, 30, 16, 8 and 3~cm. To exclude local radio interference, simultaneous observations were conducted at several widely separated locations: in the Russian Far East near the city of Ussuriisk, in the Gorky region (Staraya Pustyn'), in the Murmansk region (Tuloma) and in Crimea (Kara-Dag). In 1972, observations were also conducted aboard the research vessel ``Akademik Kurchatov'' in the equatorial waters of the Atlantic ocean \cite{Tro73a,Tro73b, Tro73c,Tro74,Bon75}. The Moscow group not only conducted searches for pulses of cosmic origin at widely separated sites, but also attempted detections of the signal dispersion that would result in later arrival of the pulses at lower frequencies with respect to high-frequency components -- the method of so-called synchronous dispersion reception \cite{Gin79,NSK77a}. The receiver consisted of a broadband low-noise amplifier covering the band 350--550~MHz and four filters at 371, 408, 458 and 535~MHz, each 5~MHz wide. Observations were carried out in 1972 in the Caucasus and the Pamir mountains. Similar observations were conducted in 1973 with receivers in the Caucasus, on the Kamchatka peninsula and aboard the Soviet Martian mission ``Mars-7''. In the latter experiment, in addition to the broad-band receiver 350--550~MHz, two lower frequency receivers at 38 and 60~MHz were used in order to match the operational bands of the ``Mars-7'' instrumentation. These observations resulted in coincidental detections of signals of several types. One of them corresponded to sporadic solar radiation, another one was related to the radio emission of artificial satellites \cite{Sal88}.

\subsection{Investigation of the statistical structure of signals}

Since the signals generated by the transmitters differ in their statistical characteristics from the noise signal of natural radio sources \cite{Go65,Sl65,Gu65,Si65}, the statistical structure of the emission of natural OH maser sources in W3, NGC 6334A, Sgr B2, W 49 and VY CMa were investigated. Observations were conducted at the Nan\c{c}ay radio telescope in France \cite{Pas71,Pas73,Lek75}. In 1972, N.S.~Kardashev and M.V.~Popov investigated the statistical properties of the radio emission of the Galactic centre at the wavelength of 3.5~cm \cite{NSK76}.

\subsection{Optimum wavelengths for interstellar communication}

A lot of attention in the first decades of SETI in the USSR was given to optimisation of the wavelength bands for interstellar communication. A detailed analysis of the problem was presented by N.S.~Kardashev who took into account fundamentally ineradicable sources of noise -- the background radiation and quantum fluctuations \cite{NSK69}. He considered two cases: 1) search for callsigns; and 2) search for informative broadcast. He assumed that in both cases, an essential part of the spectrum of the artificial source was to lie in the radio domain, and for low fluxes the signal spectrum was to be confined entirely within the radio domain from decimetre to millimetre wavelengths. Since a sender was unknown, the search should cover the entire sky. Later Kardashev considered another case in which the search was to be conducted for signals from certain celestial objects. For observations of the Galactic centre, nuclei of other galaxies and quasars the wavelength of 1.7~mm is optimal since at this wavelength the CMB has the maximum intensity. For search of signals from stars with reciprocally directed antennas, the wavelength near 1.5~mm is optimal. This band is close to the positronium line of 1.47~mm, which can be considered as a convenient benchmark, analogous in its physical significance to the atomic hydrogen line of 21~cm. Finally he came to the conclusion that for an isotropically emitting transmitter, the optimal wavelength is 21~cm, and for directed radiation -- 1.5~mm \cite{NSK79}. V.S.~Troitsky, using different arguments, also pointed at the advantage of the millimetre domain \cite{Tr86a}. The choice of band might also depend on the nature of the signal. For pulsed signals with compensating delays, the optimum frequency was found to be 56~GHz ($\lambda = 5.35$~mm) \cite{Gi73,Gi73e,Gi75r}. One more approach was proposed in 1991 by L.M.~Gindilis and V.S.~Strelnitsky \cite{Gi91,Gi93a}. Strelnitsky drew attention to the fact that the hyperfine structure of the excited level $n = 2$ of the hydrogen manifests itself by six spectral lines. All of them fall into the radio domain: three lines have frequencies around 1~GHz and three other -- near 10~GHz. Unlike the 21~cm line, these lines are not subject of interference from the galactic radio emission. In addition, use of all six lines at once gives rich possibilities for information encoding.

Since the potential interstellar communication bands need protection from other ``users'', appropriate steps have been taken in this direction. The Soviet representative in the International Telecommunication Union (ITU), B.A.~Dubinsky introduced a number of proposals for the protection of frequencies for SETI, which had been included in the Radio Regulations and other documents of the ITU Radio Communication Advisory Group (RAG) \cite{Dub74}. Despite these protective decisions, the real situation with radio frequency interference on Earth and in the near-Earth space remains unfavourable for SETI and radio astronomy in general, and there is a tendency towards its further deterioration. One of the radical solutions to this problem is placing radio telescopes on the far side of the Moon and protection of the far side from anthropogenic radio emission. In the beginning of 1970s, B.A.~Dubinsky proposed a new approach to the allocation of frequencies in the shielded area of the Moon's far side to RAG: instead of the usual allocation of separate frequency bands for various services, he suggested considering the entire radio frequencies domain in this shielded area allocated to radio astronomy and other passive radio-physical cervices, including SETI. This approach was recognised, and in 1979 the World Administrative Radio Conference included a special resolution in the Radio Regulations, which is the legal basis for recognising the shielded area of the Moon's far side as a reserve for passive radio research \cite{ITU79}.

\subsection{SETI and radio communication strategy}

The main directions of SETI radio communication strategy were formulated at the First All-Union Conference on the Search for Extraterrestrial Civilisations in 1964. One direction of searching for civilisations of our and slightly higher levels of development was focused on the search for highly directed narrow-band radiation. Another approach addressed a search for signals from super-civilisations. Subsequently, several interesting ideas were expressed by P.V.~Makovetsky \cite{Ma76,Ma77,Ma78,Ma81}. He suggested looking for callsigns at the frequencies $\pi \nu_{\rm H}$ and $\nu_{\rm H}/\pi$, where $\nu_{\rm H}$ is the frequency of the hydrogen radio line of 21~cm. In order to reduce the uncertainty in time, he suggested to consider supernova and nova outbursts as synchronisation markers. Makovetsky calculated the optimal communication periods for several nearby stars, using as a synchronisation marker the outburst of Nova Cygni 1975. RATAN-600 conducted observations of the Barnard's ``Flying'' Star in 1978 on the dates calculated by P.V.~Makovetsky but with no detection. Further on, in order to reduce the uncertainty of the direction of the search, he proposed to concentrate the attention in the direction of some objects of special importance that supposedly should be known to all civilisations of the Galaxy and can be used by them as "natural beacons" -- much in the same way as light or radio beacons in maritime  navigation.

Unlike the Makovetsky strategy, which is based on the use of ``unmodulated'' callsigns, N.T.~Petrovich considered the method of transmitting modulated signals that allow delivery of some meaningful information \cite{Pe86}. In order to avoid distortion (e.g., dispersion) of the signal in the interstellar medium, Petrovich proposed using a relative modulation methods in which information is encoded not by the absolute value of the signal parameter but by its relative value with respect to the value of the same parameter transmitted in the neighbouring time interval or on the adjacent carrier frequency. The highest immunity against noise is characteristic for phase-manipulated signals, which are used widely in the ground and space communication lines. He also considered the methodology for organising a communication channel designed to receive signals below the noise level. In this context, the problem of signal detecting via ``eavesdropping'' was analyzed by A.V.~Arkhipov \cite{Ar86}.

\subsection{Radar studies of the Lagrange points}

In 1980--81, radar studies of the vicinities of the Earth-Moon Lagrangian points L4 and L5 were carried out at the NIRFI in Gorky in search for ETI's probes \cite{Su86}. The night-time experiment was carried out at a frequency of 9.3~MHz. Radio signals were formed as pulses of a 1~s duration, separated by a gap of 4~s. The effective power was 25~MW. The reflected signal reception was carried out within a 1.5~kHz bandwidth with the integration time of 0.2~s. The duration of one session was 40~minutes. A total of about 25 sessions were conducted.

\subsection{Radio telescopes for SETI}

As noted above, one of the original tasks of the RATAN-600 radio telescope was to conduct SETI observations. Indeed, this telescope conducted several SETI observing programmes -- see sub-section \ref{sub-Zodiac} below.

In the 1980s, the NIRFI group led by V.S.~Troitsky developed a ``Survey'' system designed to search for ETI signals from an unknown direction \cite{Tr86b}. The system included several dozens of antennas with a diameter of 2~m, which together covered the whole sky. The use of multichannel receivers was envisaged. Unfortunately, this modest and inexpensive system was not implemented due to lack of funds. Great hopes were associated with the radio telescope RT-70. Its concept was developed at IKI under the leadership of N.S.~Kardashev \cite{Gi90a,Gi90b} and is based on the design of the deep space communication antenna in Evpatoria, Crimea \cite{Kozlov82}. This telescope with a 70~m parabolic reflector was designed for observations at wavelengths down to 1~mm, which would make it possible to carry out observations of the positronium line at 1.47~mm. It was also intended to use this telescope as an Earth-based component of a Space VLBI system. The construction of RT-70 started in the mountainous region of Uzbekistan on the Suffa plateau at an altitude of more than 2000~m above sea level in the 1980s. The collapse of the Soviet Union in 1991 prevented timely implementation of this project.

In 1982, the Council on Radio Astronomy of the USSR Academy of Sciences created a working group on a prospective telescope under a tentative name ``Square Kilometre Radio Telescope''. The group was led by Yu.N.~Pariiskii and involved radio astronomers and engineers from all major radio astronomy establishments of the USSR. SETI was included in the list of tasks to be addressed by the prospective facility. The working group discontinued its work in the beginning of the 1990s due to severe funding problems after the dissolution of the USSR. However, various engineering developments coordinated by this working group found their way into upgrades of the meter and decametre radio astronomy facilities in Russia and Ukraine \cite{Bovkun90}. In the 1990s, Yu.N.~Pariiskii and other members of the working group provided input into the initiative that resulted in the formation of the international Square Kilometre Array (SKA) project.

Considerable attention was given to the development of orbital radio telescopes. In the 1970s, a project of space-borne radio telescope with a diameter of more than 1~km was developed under the leadership of N.S.~Kardashev at IKI with the participation of industrial partners. The telescope was supposed to be assembled in orbit from separately transported blocks, and its surface could be built up unlimitedly \cite{Bu78}. An important milestone in the implementation of these plans was the launch of the first 10-m space radio telescope KRT-10 in June 1979 \cite{Harl05,Davi97-10}. This direction was further developed in the projects KRT-30/RAKSAS \cite{Davi97-30} and  Radioastron \cite{VVA81,VVA87}. Prospects of using space radio astronomy for SETI purposes were analysed by G.S.~Tsarevsky \cite{Ts86}.

\subsection{Search for signals in the optical domain}

Alongside various  SETI programmes in the radio domain, there were searches of optical ETI signals in the USSR too. They commenced at the Special Astrophysical Observatory (SAO) of the USSR Academy of Sciences under the leadership of V.F.~Shvartsman in the 1970s. The SAO group created a unique instrumental set that allowed them to analyse ultrafast optical variability on time intervals from $10^{-7}$ to 100 seconds. The equipment was used for both astrophysical observations and optical SETI. This activity was conducted within the framework of the programme called MANIYa (a Russian acronym for Multi-channel Analyser for Nanosecond Variations of  Brightness)  \cite{Shv77}. In the SETI context, the task was to search for ultra-narrow emission lines with the width of up to $10^{-6}$ angstroms or a pulsed laser emission. The MANIYa group composed a list of prospective ETI objects suitable for optical observations. According to Shvartsman, the most promising were radio-emitting objects with a continuum optical spectrum (ROCOS). The first observations of the MANIYa programme were carried out in 1973--74 with the telescope ``Zeiss--600'' \cite{Be77,Ev77}. Since 1978, MANIYa observations were also conducted on the SAO's 6-meter BTA telescope \cite{Shv86}.

\subsection{Defining SETI targets}

Along with the search for signals, several studies were carried out on possibilities of detecting signatures of astro-engineering activity of hypothetic ETIs (S.A.~Kaplan \cite{KapNS81}, N.S.~Kardashev \cite{NSK69,NSK81,NSK85,NSK86}, K.K.~Rebane \cite{Re86} and V.I.~Slysh \cite{VIS85}). They paved the way for SETI observing programmes carried out in the 1990s (section \ref{Ru-1990s}).

In the early 1980s, at the initiative of N.S.~Kardashev, an attempt was made to select promising SETI celestial tragets. Within the framework of this initiative, V.A.~Zakhozhay and T.V.~Ruzmaikina analysed the list of nearest stars (within 10~pc from the Sun) and selected prospective planetary system candidates \cite{ZaRu86}. Today, after discovering many extrasolar planets, this work might be seen as obsolete, but in the 1980s the topic was highly relevant. V.G.~Surdin demonstrated a different approach. He examined the conditions in globular clusters and showed that the stars of globular clusters may have terrestrial-type planets. Since the distances between stars in globular clusters are not large, intelligent beings living on these planets could easily establish radio communication among themselves. Surdin selected candidates from the catalog of globular clusters to search (intercept) interstellar communication signals \cite{Su85,Mi07}. Selection of potential SETI targets continued to be at the centre of attention in the following years (e.g., \cite{Pan14}).

\section{Theoretical studies}
\label{Theor-std}

\subsection{Discussion on the plurality of inhabited worlds, the Fermi paradox}

In 1975, at the Zelenchukskaya school--workshop on CETI, I.S.~Shklovsky presented a concept of the uniqueness of our Earth civilisation \cite{ISSh76}. This caused a discussion, in which N.S.~Kardashev, V.S.~Troitsky and others criticised the concept \cite{NSK77b,Mak79}. A debate on this issue took place between I.S~Shklovsky and S.~Lem \cite{Lem77}. These discussions indicated the need for a more rigorous approach to assessing the number of communicative civilisations. L.S.~Marochnik and L.M.~Mukhin estimated the number of civilisations in the Galaxy, based on the concept of life emerging within a narrow ring zone of the Galaxy near the co-rotation area, and the lifetime of civilisations is determined by the time of the stars' movement along the galactic orbit between adjacent spiral arms \cite{MaMu83,MaMu86}. L.M.~Gindilis applied a statistical method for estimating the number of civilisations \cite{Gi81}. In a general and most rigorous form this problem was later addressed by A.D.~Panov \cite{Pan07}. The problem of the Galactic population was also considered by V.S.~Troitsky \cite{Tr81y}. He put forward a new concept of the simultaneous one-off act of nascency of life in the Universe. Based on this hypothesis, Troitsky developed a theory of the Galactic population, which differs from the conventional theory. He demonstrated that for a finite lifetime of civilisations, their number does not remain constant over time, as follows from the Drake's formula, but tends asymptotically to zero.

The concept of uniqueness of the Earth civilisation is largely based on the absence of a ``cosmic miracle'' (the Fermi paradox). In the framework of addressing the apparent uniqueness, V.S.~Troitsky analysed the possibility of creating powerful omnidirectional beacons-transmitters for interstellar communication and came to the conclusion that the need to preserve the circumstellar habitat leads to energy constraints that would not allow operating of a sufficiently powerful transmitter corresponding to the Kardashev's types II and III civilisations  \cite{Tr86a}. A detailed analysis of the problem of the multiplicity of inhabited worlds, including the AS paradox\footnote{AS (astrosociological) paradox -- a contradiction between the idea of a multiplicity of extraterrestrial civilisations (based on the Copernicus--Bruno principle) and the lack of observational manifestations of their activity in outer space (radio signals, etc.). In its strongest form, the AS paradox is interpreted as a contradiction between the assumed multiplicity of ETI and the absence of traces of their presence on Earth -- the Fermi paradox.}, was carried out in 1988 by L.M.~Gindilis \cite{Gi88} and reviewed by Yu.N.~Efremov \cite{Efremov11}.

\subsection{Models of development of cosmic civilisations}

Two strategies of SETI, formulated at the first All-Union Conference on Extraterrestrial civilisations (Byurakan, 1964), were based on two different concepts of the development of ETIs. One of them proceeds from the premise that the power level of civilisations is limited by certain physical and ecological reasons (V.S.~Troitsky). An alternative approach assumes the possibility of achieving a much higher power level, up to $10^{38}$~W, comparable to the power release of a whole galaxy (N.S.~Kardashev). According to Kardashev, civilisations should strive for unification in compact systems in order to collect all of their resources in a relatively small number of objects (the hypothesis of ``urbanisation''). He examined six different scenarios for the development of civilisations, in which unification occurs at different spatial scales \cite{NSK85,NSK86}. A more general approach based on a system analysis is presented in the works by L.V.~Leskov \cite{Les83,Les86, Les85}.

\subsection{The contact problems} 
\label{Contacts-sub}

The most important aspect of contact between civilisations is the possibility of mutual understanding under different systems of semantic concepts. In the 1970s, this problem was actively developed by B.N.~Panovkin. He came to the conclusion that a contact between civilisations through communication channels is impossible \cite{PBN71,PBN76}. The discussion on this topic \cite{Gin73z,Kaz97}, in which V.V.~Kazyutinsky took part, proved to be very useful, because it contributed to a more rigorous assessment of the real state of the problem. Another direction of studies was developed at the Institute of Cybernetics of the Ukrainian Academy of Sciences under the direction of I.M.~Krein. She considered the contact problem in the context of developing ``languages--mediators'' \cite{Kre81,Kre86,Chu86}. The problem of languages for communication with cosmic civilisations was considered by B.V.~Sukhotin. He did not set the goal of constructing a special language for communication with ETIs, but solved the problem of deciphering a message received through interstellar communication channels, provided that it was ``written'' with unknown symbols in an unknown language \cite{Su69}. A thorough philosophical analysis of the problem of contact with ETIs was made by V.V.~Rubtsov and A.D.~Ursul \cite{RuUr87}.

A constructive and overall ``pro-contact'' opinion on the SETI problem in general and the issue of communication between cosmic civilisations in particular was expressed by A.D.~Sakharov in the answers to the questionnaire distributed among 64 scientists prior to the 1971 US-USSR CETI Conference in Byurakan (published in \cite{Gin90, Gi93s}). 

\subsection{Interstellar flights}

The theory of relativistic celestial mechanics was developed in the USSR in the 1970s and 1980s with a comprehensive contribution by V.A. Brumberg \cite{Brum72,Brum91}. Its applications covered interstellar relativistic flight dynamics, including issues related to the search of extraterrestrial civilisations \cite{Zak84,MaZa86}. V.G.~Surdin considered the problem of traveling within the Galaxy using single and multiple gravitational maneuvering around stars of various types. He came to the conclusion that the best conditions for such a gravity assist interstellar travel exist in the nuclei of globular clusters \cite{Su89}.

\subsection{Philosophy and SETI}

Studies of the possibilities of communication with extraterrestrial civilisations lead to the formulation of a number of questions of general scientific and philosophical interest. The development of philosophical aspects of SETI in the USSR involved V.V.~Kazyutinsky, V.V.~Rubtsov, A.D.~Ursul and others. Their works were regularly presented and discussed at the annual "Kaluga Readings", devoted to the development of the scientific heritage and ideas of the pioneer of exploration of space Konstantin Tsiolkovsky and his younger follower and fellow countryman Alexander Chizhevsky, both prominent figures in the development of the philosophical school commonly dubbed as ``Russian Cosmism''.  The venue of the event was not arbitrary: Tsiolkovsky and Chizhevsky lived and met in the first quarter of 20th century in Kaluga, a provincial city some 190 km south-west of Moscow. In a way, theoretical works presented at the ``Kaluga Readings" in the 1960s-80s were further developments of the ideas of the Kaluga cosmic philosophers, summarised in the books by Tsiolkovsky \cite{Tsiolk01} and Chizhevsky \cite{Chizh04}. The materials of the Kaluga Readings are published in the proceedings reviewed in \cite{Komarov86} and references therein. A comprehensive introduction to Russian Cosmism with multiple links to the philosophical issues relevant to SETI were published recently by G.M. Young \cite{Young12}.

Several fruitful discussions on the philosophical aspects of the SETI problem hosted at the Institute of Philosophy of the USSR Academy of Sciences were included in a special book \cite{AsMeId}. 

A new angle to the SETI problem was suggested by V.F.~Shvartsman. He argued that the problem of SETI is not only an astrophysical and even not only a general scientific problem, but a problem of the entire human culture \cite{VFSh86}.  This paper attracted attention of a very broad audience, well beyond the traditional circles of researchers in the SETI field.

\section{SETI in Russia in the 1990s -- 2000s}
\label{Ru-1990s}

The general decline of science in Russia after the collapse of the Soviet Union could not but affect the state of SETI. However, owing to the enthusiasm of the researchers, SETI studies have not ceased all together. Experimental research has evolved in several directions: (1) search for radio signals from Sun-like stars; (2) search for optical ETI signals; (3) search for Dyson spheres; and (4) transmission of radio messages to extraterrestrial civilisations. In addition, a number of researchers continued theoretical studies.

\subsection{Search of radio signals and radio communication strategy for SETI, the ``Zodiac'' programme}
\label{sub-Zodiac}

The ``Zodiac'' programme of searching Sun-like stars was initiated by L.N.~Filippova and V.S.~Strelnitsky. The first observations of this program were carried out in October 1989 using the radio telescope RATAN--600 and continued in subsequent years under the initiative by L.N.~Filippova. A.M.~Baryshev, N.N.~Bursov, O.V.~Verkhodanov, I.V.~Gosachinskiy, M.G.~Mingaliev, V.N.~Sidorenkov, V.A.~Stolyarov and others took part in the observations and processing of the RATAN--600 ``Zodiac'' data. For this programme, 29 stars were selected from the D.~Soderblom's list \cite{Sod86}, all located within 14 degrees from the Ecliptic, and several solar-type stars closest to us. Subsequently, a broader list of ``SETI candidates'' stars was composed. In total, over the period 1994--2006, 47 stars of the ``SETI candidates" list were observed, including 5 stars with planetary systems. Observations with RATAN--600 were carried out in two modes, meridian transit and tracking. In the transit mode, the observations were conducted at 1.0, 1.38, 2.7, 3.9, 6.25, 7.6, 13 and 31~cm wavelengths simultaneously. In the tracking mode, observations were made at 21~cm. Some noteworthy features were observed in the star W 252 \cite{BuMi07, Fi12}. Several stars were also observed in the optics using the SAO's 6-meter BTA telescope. In 2002--2005, G.M.~Rudnitskii searched for narrow-band signals from the nearest stars (mainly of the solar type) using the RT-22 radio telescope in Pushchino. These observations were conducted at 1.35~cm (the water maser line) and 8.2~mm (the cyanacetylene line) \cite{Rud07}.

In 1998, under the initiative by S.F.~Likhachev, four stars from the L.N.~Filippova list were included in the VLBI program of the INTAS-98 experiment involving 6 radio telescopes. In the direction of one of the stars, 37 Gem, a point source was detected at the frequency band 1664.99--1666.95~MHz with the intensity twice exceeding the noise level. The nature of this emission is still unclear \cite{Chu07}.

A group at the Astro Space Center (ASC) of the Lebedev Physical Institute (FIAN) continued searching for Dyson spheres. They focused on IR-objects from the IRAS catalogue and selected 98 prospective Dyson sphere candidates, of which 40 objects could be identified with astronomical objects of various types; 58 objects remained unidentified \cite{Ti00}.

N.S.~Petrovich continued developing a strategy for finding signals immersed in strong noise \cite{Pet95a,Pet95b,Pet97}. He stressed that detecting a signal below the noise level permits a scheme for constructing a galactic communication, based on low-gain antennas instead of high-gain ones. Such a scheme would make possible serving many potential recipients at once and thereby increasing the probability of establishing a contact.

P.A.~Fridman investigated the data rate of potential radio communication channels between cosmic civilisations in the presence of noise \cite{PAF11} and analysed the applicability of his algorithm for detecting radio transients \cite{PAF10} in the SETI context. 

\subsection{Search for optical signals: continued}

In the 1990s--2000s, the programme of searching for optical ETI signals continued under the leadership of G.M.~Beskin (SAO). The list of objects for optical SETI has been expanded to include F9V--G5V spectral class stars within a distance of 25~pc (for searching for type I civilisations) and objects with unusual characteristics (ROCOS and white dwarfs of the DC-type) for super-civilisations search.

Approximately 20 objects of each type were observed, and an upper limit on the power of hypothetical ETI lasers was estimated. In the early 1990s, a set of MANIYa equipment was installed on the 2-meter CASLEO telescope in Argentina, enabling observations of the southern sky objects \cite{Bes98,Bes07}.

A group of scientists from FIAN and Scientific--Production Association ``Astrofizika'' developed a laser receiver with a quantum sensitivity limit at 1.315~$\mu$m for the atomic iodine laser transition $^{2}P_{1/2} \to \, ^{2}P_{3/2}$ capable of detecting laser signals against the background of a star selected for observation. This wavelength, fixed in the spectrum, is, in the opinion of the authors, useful as a spectral reference for the optical SETI \cite{Kut07,Kut10}.

\subsection{Transmission of radio messages to extraterrestrial civilisations, METI}

Until the late 1990s, studies in the field of SETI in the USSR and Russia were limited to searching signals per se. However, in the following years, several projects of active messaging, including international ones, were conducted (Table~\ref{METI-table}). All messages were sent from the Evpatoria Planetary Radar in Crimea, Ukraine. Much credit for this activity goes to A.L.~Zaitsev, under whose guidance all these experiments were performed by an international team that has developed, among other things, suggestions on the targets and contents of messages (e.g. \cite{1998AAS...193.9710D}.)  Zaitsev developed the principles of organising the message signal. The message should contain three components: (1) a monochromatic signal with a constant frequency, freed from Doppler drift caused by rotation of the Earth around its axis and its orbital motion around the Sun; (2) a discrete (in the simplest case -- binary) frequency-modulated signal, which carries logical information (knowledge transfer); (3) an analog signal carrying music (emotion transfer). Technically, for the transfer of music, Zaitsev suggested using a non-contact musical instrument called thereminvox, which generates narrow-band quasi-sinusoidal oscillations with frequency modulation, described by a smooth single-valued function with a continuous phase. In 2000, he submitted a proposal to the National Astronomical and Ionospheric Center for transmitting a thereminvox music using the Arecibo Telescope in active (radar) mode toward potential ETI recipients. This idea was implemented in 2001 in the Teenage Message project, the signal of which contained all three components listed above \cite{Gin02}. In response to the criticism of the METI activities, A.L.~Zaitsev justified the necessity of transmitting messages  in a number of works \cite{Za04}. In particular, he presented arguments against the alleged danger of signal broadcasting. This position was supported by L.M.~Gindilis \cite{Gin11}. It should be noted that Russian scientists who discussed the problem of SETI at the conference ``Horizons of Astronomy and SETI'' (Special Astrophysical Observatory of the Russian Academy of Sciences, September 2005) came to the conclusion that active signalling to ETIs is necessary in parallel with searches for their signals. This was reflected in the memorandum of the conference \cite{Mem07}.

\begin{table}[h]
\caption{METI campaigns at the Evpatoria Deep Space Communication Centre}
\label{METI-table}
\begin{tabular}[h]{|l|c|c|c|c|}
\hline \hline
Campaign & Cosmic Call 		& Teenage 	& Cosmic Call 		& A Message 	\\ 
name	& 1999               	& Message	& 2003           		& from Earth	\\ \hline
Dates	&  24.05, 30.06, 	& 29.08, 03.09  & 06.07.2003 		& 09.10.2008 	\\
        		& 01.07.1999     	& 04.09.2001    &                    		&                    	\\ \hline
Team	& Chafer, Dutil,	 	& Pshenichner, & Chafer, Dutil, 		& Madgett, Coombs, \\
		& Dumas, Braastad,	& Filippova, 	& Dumas, Braastad,	& Levine, Cooper, \\
		& Zaitsev, et al.		& Gindilis,		& Zaitsev, et al.		& Zaitsev, et al.	\\
		&				& Zaitsev, et al.	&				&			\\ \hline
Number 	&	4			&	6		&	5			&	1		\\
of sessions &				&			&				&			\\ \hline
T [min.]	&	960			&	366		&	900			&	240		\\ \hline
E [MJ]	&	8640			&	2200		&	8100			&	1440		\\ \hline
Reference &	\cite{Za99}	&  \cite{Gin02}	& 	\cite{Za03}	& \cite{Kiss08}	\\ \hline
\hline
\end{tabular}
\end{table}

\subsection{The origin of life in the Universe: the Galaxy population theory}

 Generally accepted ideas on the origin and development of civilisations are based on the assumption that civilisations emerge continuously. V.S.~Troitsky renounced the idea of the continuous nascency of life in the Universe and suggested that life has been born only once and simultaneously in the entire Universe, during a narrow interval of time of the cosmological evolution, and it has happened on those planets where by that time the necessary physicochemical conditions have been favourable \cite{Tr81y,Tr86a}. He gave a detailed substantiation of this hypothesis in his last work, published after his death \cite{Tr96}.

A.D.~Panov considered origins of life by means of panspermia of prebiotic evolution products \cite{Pa07,Pa14}. His quantitative model leads to the conclusion of an increase of the probability of the emergence of life by many orders of magnitude comparing to the model assuming a prebiotic evolution on an isolated planet. In the Panov's model, life emerges almost simultaneously on many planets, on which the necessary conditions are ripe, with the same molecular basis, a single genetic code and with one chirality.

\subsection{A cosmic subject of Lefevre, a fast burster and black holes}

V.A.~Lefebvre developed a mathematical model of a reasonable cosmic subject, whose distinguishing feature is the presence of conscience, and considered the astronomical applications of this model \cite{Lef96}. Later astronomical aspects of this model were considered in the article by V.A.~Lefebvre and Yu.N.~Efremov \cite{Lef00}. The authors proceeded from the proposition that the search for cosmic civilisations will acquire the status of a strictly scientific task if it is possible to create a theoretical model of the world that contains a rational subject as a natural component. Such a model should relate the phenomenon of intelligence to the physical picture of the Universe and provide indications of possible observable signs of artificial activity. They drew attention to the X-ray source MXB~1730-335, the so-called Fast Burster (FB) that manifests emission irregularities consistent with the Lefebvre model. The model also offers important interpretational applications for black holes. Lefebvre and Efremov attracted attention to the amazing parallel between the inner world of a Kerr's black hole and a psychological model of a reflecting (thinking) subject.

\subsection{Cosmology and SETI}

N.S.~Kardashev has  been investigating cosmological aspects of SETI \cite{NSK97,NSK99,NSK02}. This includes considerations on possible ways of ETI's evolution and the consequent strategy of their search for wormholes and mirror civilisations. It is believed that the exchange of information between our and a mirror world is possible only gravitationally. However, Karadashev indicated another possibility associated with black holes. The Hawking radiation has three components: electromagnetic, gravitational and corpuscular. In the presence of mirror matter, the intensity of Hawking radiation doubles. If the Hawking radiation can be controlled by changing the mass of a black hole (by changing the accretion rate), it might be used for transmitting information by means of modulated electromagnetic radiation.

\subsection{The Higher Intelligence in the Universe, a Supermind}

In the mid-1990s, V.M.~Lipunov published an article entitled ``Scientifically Discovered God'' \cite{Lip95}. He draws attention to the fact that, contrary to the existing prejudice, there is nothing non-scientific in assuming the existence of the Supermind. Nature, which has the opportunity to trigger emergence of life over an almost infinitely long period of time, sooner or later should produce a Supermind. Discussing Einstein's ideas on the cognizability of the World, Lipunov emphasised the impossibility of simultaneous recognition of the infinite complexity of the World and its successful cognition, while not recognising the existence of the Supermind. G.M.~Idlis came to the conclusion about the existence of the Higher Intelligence based on the mathematical analysis of the organisation of matter at the physical, physicochemical, chemical-biological and psychological levels \cite{Kuz96}. The discussion of the problem of the Higher Intelligence was largely stimulated by the considerations of the anthropic principle and the astrosociological paradox. The latter has been addressed in the papers by Gindilis \cite{Gi96}, Gindilis and Rudnitskii \cite{Gi93b}, and Yazev \cite{Yaz98}.

\subsection{Pedagogy of SETI}

Many sides of the SETI problem proved to be very attractive for schoolchildren and students. Educational aspects of SETI got considerable attention in the USSR and Russia \cite{Gin72,Per98,Fil00,Lev00,Feo01,Ten07,Dmi07,Lev07,Gin08}. In particular, SETI-related lectures and laboratory exercises were organized at the All-Russia Pioneer Camp ``Orlyonok'' near Tuapse by L.N.~Filippova and at the Moscow City Pioneers' Palace (now the Palace of Children and Youth Creativity) by B.G.~Pshenichner, I.A.~Feodulova and N.V.~Dmitrieva.

\subsection{SETI in the general astronomical context}

The SETI topics were discussed regularly at the All-Russia Astronomical Conferences, at the conferences of the Astronomical Society and at the sessions of the A.S.~Popov Scientific and Technological Society of Radio Engineering, Electronics and Communication (NTORES). Worth noticing are several conferences dedicated specifically to SETI: Scientific and Methodological Conference ``SETI: the past, present and future of civilisations'', Moscow, 24--27 May 1999; Scientific Conference ``SETI at the threshold of the 21st century: results and prospects'', Moscow, 5-7 February 2002; Scientific Conference ``Horizons of astronomy and SETI'', Nizhny Arkhyz, 25--30 September 2005; the 3rd Symposium of the International Academy of Astronautics ``Search for signs of life'', St. Petersburg, 27--29 June 2011 \cite{IPA2011}.

\section{SETI in Ukraine}
\label{Ukr-1990s}

In Ukraine, the problem of SETI during Soviet and post-Soviet times has been addressed at several institutes of the National Academy of Sciences of Ukraine (NANU). In the 1960s-80s, as mentioned in sub-section \ref{Contacts-sub}, I.M. Krein and her colleagues at the V.M.~Glushkov Institute of Cybernetics in Kyiv investigated interactions of a human with intelligent and highly organised systems \cite{Krein86}.

A number of original ideas were put forward by A.V.~Arkhipov at the Radio Astronomy Institute in Kharkiv. In 1986, he considered the possibility of detecting leakage signals similar to Earth's television broadcast and identified 4 candidate sources that met the criteria for such the signals \cite{Ar86}. Later, he proposed an original strategy for finding ETI signals. It is based on the assumption that in order to protect astro-engineering structures from the ionising radiation of their star, a civilisation can create around them an artificial magnetosphere. Interaction of this magnetosphere with the interplanetary plasma should lead to the generation of non-thermal cyclotron radio emission in the domain of decametre waves. Detection of such the radio emission could serve as an indication on the existence of artificial magnetospheres. After analysing the data from the sky survey conducted by the decametric radio telescope UTR-2 and comparing them with the catalog of nearby stars, Arkhipov singled out the source GR~0752$-$01, which coincides with the star HD~64606 of the spectral class G8V located at a distance of 19~pc from the Sun. It can be considered as a possible candidate for SETI. Another SETI approach suggests intercepting of radio communications of an ETI probe positioned in the Solar System. A typical approach to searching informative signals assumes that the signal is addressed to our civilisation. Arkhipov considered a more realistic problem -- an interception of radio transmission of a hypothetic radar of the ETI's probe or communication signals between this probe and its parent civilisation.

Along with the radio communication strategy, A.V.~Arkhipov developed an ``unconventional'' search strategy for the ETI associated with potential discoveries of non-terrestrial artefacts on Earth and Moon. He applied this idea in two forms. The first is a search of artefacts associated with possible reconnaissance missions of ETI in the Solar System. The second focuses on a search for a ``waste'' of the ETI probe's activities (e.g., space debris). He argued that the best conditions for searching for artefacts of the first type were realised on the Moon. Arkhipov showed a principle feasibility of discovering the ETI artefacts on the Moon and formulated the principles of Lunar archeology, which might be of interest not only for SETI, but also for Lunar exploration projects. He also estimated that from 3\% to 15\% of space debris are thrown into the interstellar medium and can reach the habitat of another civilisation. Having estimated the frequency of the ``artificial non-terrestrial debris'' hitting the Earth's atmosphere and possible survival of atmosphere descent, he concluded that some debris could reach the Earth's surface \cite{Arkhipov96}. He also argued that such contamination of Earth by artefacts from about $10^{6}$ nearby stars could reach Earth as agents of spontaneous interstellar panspermia (\cite{Arkhipov97} and references therein). In this respect, Arkhipov drew attention to the need of studying the so-called ``pseudo-meteorites'' and ``fossil artefacts''.

The possibility of contamination of the Earth (and other planets) with waste from ETI's space activities offers a new angle to the study of the problem of panspermia. Due to the proliferation of debris around each ``technogenic'' star, there is a non-sterile zone, which includes microorganisms. When the Solar System moves through the Galaxy, it crosses the non-sterile zones of various stars, while non-sterile artefacts can get to Earth. The same would happen to other planetary systems. According to Arkhipov, just  0.7\% of the rate of production by our civilisation is sufficient for infecting an Earth-like planet. About $10^{5}$ stars could infect Earth during its lifetime. These studies, carried out mainly in the 1990s, were published by A.V.~Arkhipov in numerous articles, both in Russian and foreign journals. Their generalisation is contained in his Ph.D. thesis \cite{Ar98}.

In recent years, A.V.~Arkhipov put forward original ideas on possible detection of artificial structures in the framework of studies of variability of stars and search for exoplanets \cite{Ar07,Ar12}. A brief review of the SETI research in Ukraine is presented in the work by L.N.~Litvinenko and A.V.~Arkhipov \cite{Lit93}.

\section{SETI prospects}
\label{Cnclsns}

The prospects of SETI, in our opinion, are defined, first of all, by the developments in astronomical observing techniques. In the coming years, new ultra-sensitive instruments capable of obtaining and processing huge amounts of observational data will become operational. {The most exciting perspectives in radio domain are offered by the Square Kilometre Array (SKA) \cite{Siemion14}.}The efforts in the field of SETI will be aiming at covering up the entire range of electromagnetic waves - from long-wave radio to gamma domains. It is very likely that attempts will be made to use signals of a different nature, most likely -- gravitational waves and neutrinos, as the technique of their detection and generation is being rapidly developed. One cannot rule out the possibility of the appearance of entirely new ``information channels'' based on the laws of nature that are still unknown to us, which manifest themselves in still unrevealed forms of matter. This is a distant prospect. However, at present, we are witnessing how our ideas on life in space change under the influence of astronomical discoveries. In the middle of the twentieth century, the only conceivable way of detecting life outside the Solar System was via receiving intelligent signals. By the end of the 20th century other possibilities for searching for Cosmic life have emerged. In recent years, the extraordinary adaptability of life to the most diverse, including extreme, conditions has been revealed. This leads to a new approach to investigating the possibility of life on the planets of the Solar System, their satellites, asteroids, comets, in the interplanetary and interstellar media. Of great importance is the discovery of signs of life in meteorites. There is increasing evidence that everywhere in the Universe (and certainly in our Galaxy) molecules of fairly complex organic compounds are common. These molecules survive the phase of planet accumulations from gas-dust disks and can be involved in composing meteorites and cometary nuclei. The latter can deliver them to the emerging or evolved planets. The avalanche of exoplanet discoveries over the past 20 years has also changed the strategy of SETI. While early SETI attempts were largely blind, now it is possible to focus attention on stars surrounded by planets with physical conditions favourable for the existence of life. This creates a stage for experimental astrobiology. The overall range of issues belonging to the realm of SETI's interests has expanded enormously over the past half a century.

Astronomy and other branches of natural, engineering and humanitarian sciences undergo very rapid development over the past decades. One might expect that we will encounter completely unpredicted discoveries, which will fundamentally change our ideas on the SETI problem. \\

\bigskip

{\it Acknowledgements.} The authors express gratitude to the Acta Astronautica reviewers and Rebecca Charbonneau for careful reading of the manuscript and many constructive suggestions for its improvement. \\

\bigskip
 

{\bf Appendix: List of acronyms}

\begin{tabbing}
MANIYa****  \= last text  \kill
APP     \>  Antenna of Variable Profile                     \\
ASC    \> Astro Space Center of P.N.~Lebedev Physical Institute \\
BTA    \> Large Azimuth Telescope (the 6-m optical telescope of SAO)  \\
CASLEO \>  El Leoncito Astronomical Complex, Argentina     \\
CETI    \> Communication with Extraterrestrial Intelligence \\
CMB \> Cosmic Microwave Background \\
ETI      \> Extraterrestrial Intelligence \\
FB   \> Fast Burster \\
FIAN   \> Lebedev Physical Institute of the Soviet (after 1991 -- Russian) Academy of Sciences \\
GAISh \> Sternberg Astronomical Institute \\
IKI       \> Space Research Institute of the USSR (after 1991 -- Russian) Academy of Sciences \\ 
INTAS-98 \> European Commission International Association for the promotion of cooperation \\
                \> with scientists from the independent states of the former Soviet Union, the project of 1998 \\
KRT-10 \> Space Radio Telescope with 10-m antenna \\
KRT-30 \> Space Radio Telescope with 30-m antenna \\
IRAS \> Infrared Astronomical Satellite \\
ITU      \>  International Telecommunication Union  \\
MANIYa \> Multi-channel Analysis of Nanosecond Variations of Brightness \\
METI \> Messaging to Extraterrestrial Intelligence \\
NANU  \> National Academy of Sciences of Ukraine \\
NIRFI \>  Radio Physical Research Institute  \\
NRAO \> National Radio Astronomy Observatory \\
NTORES \> Scientific and Technological Society of Radio Engineering, Electronics and Communication \\
RACSAS \> Radio Astronomical Space-borne Aperture Synthesis System \\
RAG \>   ITU Radio Communication Advisory Group    \\
RATAN-600 \> Radio Telescope of the Academy of Sciences with diameter 600 m \\
ROCOS \> Radio-emmitting Object with Continuum Optical Spectrum \\
RT-MSU \>     Radio Telescope of Moscow State University \\
RT-22 \> Radio Telescope with 22-m antenna \\
RT-70 \> Radio Telescope with 70-m antenna \\
SAO   \> Special Astrophysical Observatory of the USSR (after 1991 -- Russian) Academy of Sciences \\
SETI \> Search for Extraterrestrial Intelligence \\
SKA \> Square Kilometre Array (radio telescope) \\
TASS \> Telegraph Agency of the Soviet Union \\
UTR-2 \> Ukrainian T-shaped Radio telescope, modification 2 \\
VLBI \> Very Long Baseline Interferometry \\
 \end{tabbing}

\bibliography{./mybibfile-lg}

\end{document}